\documentclass{article}
\usepackage{nips2003e,times,amsmath,amssymb,amsfonts,pifont,makeidx,epsfig}

\title{Diversity and Adaptation\\
in Large Population Games}

\author{
K. Y. Michael Wong, S. W. Lim and Peixun Luo\\
Department of Physics, Hong Kong University of Science and Technology,\\
Clear Water Bay, Hong Kong, China.\\
\texttt{\{phkywong, swlim, physlpx\}@ust.hk}
}

\begin{document}

\maketitle

\begin{abstract}
We consider a version of large population games 
whose players compete for resources
using strategies with adaptable preferences. 
The system efficiency is measured by the variance of the decisions. 
In the regime where the system can be plagued 
by the maladaptive behavior of the players, 
we find that {\it diversity} among the players 
improves the system efficiency, 
though it slows the convergence to the steady state.
Diversity causes a mild spread of resources at the transient state,
but reduces the uneven distribution of resources in the steady state.
\end{abstract}

\section{Introduction}

In recent years, there is an increasing interest
in studying the adaptive behavior of players
when they learn to play games together \cite{fudenberg}.
Much work focused on learning algorithms computing Nash equilibria
for players connected by graphs \cite{kearns},
or mutually interacting through global functions \cite{kearns2}.
Since these processes are dynamical in nature,
it would be interesting to consider
how the transient and steady states of the system depend
on the choice of initial conditions.
When it reaches the steady state,
it is possible that the system evolves periodically or even chaotically,
or gets trapped in suboptimal attractors.
It is therefore important to study the collective dynamical behavior
of multi-agent systems.

In this paper, we consider the dynamics
of a version of large population games
which models the collective behavior of players 
simultaneously and adaptively competing for limited resources.
The game is a variant of the Minority Game (MG),
in which the players making the minority decision
are the winners \cite{zhang}.
It may be applied to various adaptive systems,
such as predators searching for hunting grounds with fewer competitors,
network routers trying to find the path with least delay,
or traders in a financial market trying to buy stocks at a low price
when most other traders are trying to sell (or vice versa).

We are interested in whether the entire system performs efficiently.
For example, if the MG corresponds to
the distribution of resources among a network of load balancers,
then the system is said to be efficient
if the load is uniformly distributed among the players.
Similarly, if the MG corresponds to
trading agents in a stock market,
then the system is said to be efficient
if the fractions of winners and losers are the same.

Previous work \cite{savit} showed that
when the complexity of the players' strategies is too high,
the players cannot collectively explore the strategy space thoroughly,
thus limiting the market efficiency.
On the other hand,
when the complexity of the players' strategies is too low,
the market efficiency suffers from
the {\it maladaptive} behavior of the players,
meaning that they prematurely rush to adapt to market changes in bursts.
Maladpatation is a common but undesirable phenomenon
in many adaptive systems.
As will be shown in this paper,
the introduction of diversity to the preference of strategies of the players
can lead to a better system efficiency.

\section{The Minority Game}
\label{section:MG}

The Minority Game (MG) involves a population of players 
repeatedly competing to be in the minority group
in an environment of limited resources \cite{zhang}.
Each of the $N$ players
can make a decision 1 or 0 at each time step,
$N$ being odd.
The decisions may represent buy (1) or sell (0)
if the MG models a market,
or the choice of one of two tasks
if the MG models a group of load balancers.
Each agent makes her decision independently
according to her own finite set of ``strategies''
which will be defined later.
After all players have made their choices,
players who have made decision 1 are declared winners
if there are fewer 1's than 0's,
and we denote the outcome by 1;
otherwise, players who have made decision 0 win and the outcome is 0.
The wealth acquired by an individual player
is measured by her real points,
which increases (decreases) by 1 if she wins (loses) at a time step.
The time series of 1's and 0's is called ``history'',
and is made available to all players as the only global information
for their next choices.

At each time step, the players make their decisions
based on the most recent $m$ bits in the history,
hence $m$ is known as the memory size.
There are $D\equiv 2^m$ possible histories,
thus $D$ is the dimension of the strategy space.
A strategy is then a function which maps each of the $D$ histories
to decisions 1 or 0.
Before the game starts,
each agent randomly picks $s$ strategies from the pool of strategies,
with repetitions allowed.
Each agent holds her strategy set throughout the whole game.
At each time step, the players choose,
out of the $s$ strategies she has,
the one which has so far adapted most successfully to the game,
and make decisions accordingly.

The success of a strategy is measured by its virtual point,
which increases (decreases) by 1
if it indicates the winning (losing) decision at a time step,
irrespective of whether it is chosen at that time step by an agent or not.
The availability of multiple strategies
provides an agent with adaptivity,
such that she may use an alternative strategy
when the one she chooses does not work well.
Though we only consider random strategies instead of organized ones,
we expect that the model is sufficient to capture the main features
of the macroscopic collective behavior
of many players with diverse strategies.

To model diversity among the players,
the players may enter the game with diverse preferences of their strategies.
This is done by randomly assigning $R$ virtual points  
to the $s$ strategies of each agent before the game starts,
$R$ being an odd integer.
Hence the initial virtual point of each strategy
obeys a multinomial distribution with mean $R/s$ and variance $R(s-1)/s^2$.
$R$ can thus be considered as a parameter of randomness.
The ratio $\rho\equiv R/N$ is referred to as the {\it diversity}.
Furthermore, the game is deterministic for odd $R$ and $s=2$,
since in this case no two strategies have the same virtual points throughout the game.

This is in contrast with previous versions of the game,
in which the virtual points of all strategies are initialized to zero,
corresponding to the special case of $R = 0$.
The homogeneous initial condition leads to the further simplification that
the virtual points of a strategy appears the same to all players subsequently.
However, this assumption needs to be re-examined for two reasons.
First, if the MG is used as a model of distributed load balancing,
it becomes natural to investigate
whether the artificially created maladaptive behavior
has prevented an efficient exploration of the phase space,  
thus hindering the attainment of optimal system efficiency.
Second, if the MG is used as a model of financial markets,
it is not natural to expect that all agents 
have the same preference of a strategy at all instants of the game.

\section{Main Features of the MG}
\label{section:MG result}

\begin{figure}[ht]
\centerline{\epsfig{figure=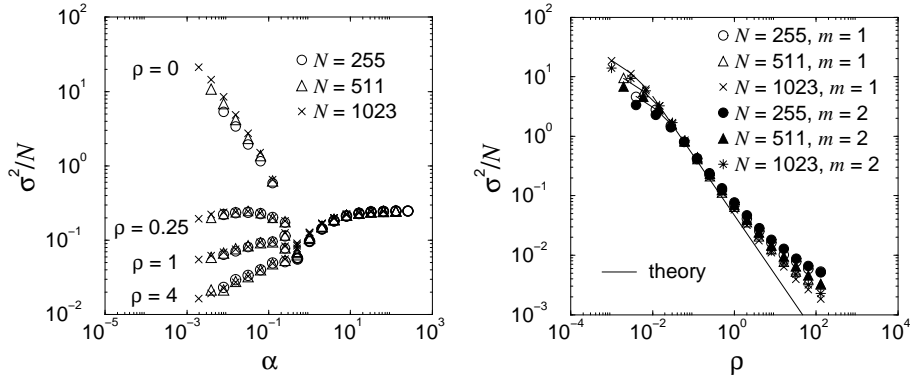,width=0.95\linewidth}}
\caption{\small (a) The dependence of the variance on
the complexity at $s=2$ and averaged over 128 samples for each data point.
(b)  The dependence of the variance on the diversity at $s=2$ 
and averaged over 1024 samples. 
Solid line: theoretical results for $m$ not too large,
with the three leftmost data points replaced by results 
for $m=1$ and small $R$.
}
\label{sd}
\end{figure}

Let $N_1(t)$ be the population of players making decision 1
at the $t$-th time step.
To study whether the game distributes resources efficiently,
we first consider the case of small $R$.
As shown in Fig. \ref{sd}(a),
the variance $\sigma^2/N$ of the population for decision 1
scales as a function of the {\it complexity} $\alpha\equiv D/N$, 
agreeing with previous observations \cite{savit}.
When $\alpha$ is small, games with increasing complexity 
create time series of decreasing fluctuations.
However, the variance does not decrease with $\alpha$ monotonically.
Instead, there is a minimum around $\alpha_c\simeq0.5$,
after which it increases gradually to a so-called coin-toss limit,
as if they were making their decisions randomly,
with $\sigma^2/N=0.25$.

The existence of a minimum variance lower than the coin-toss limit
shows that the players are able to cooperate
to improve the system efficiency,
despite the fact that the players are selfish 
and making independent decisions.
Savit $et$ $al$ \cite{savit} identified that
as the complexity changes across a critical value,
the system undergoes a phase transition.
In the high complexity phase,
the players cannot coordinate to explore the strategy space effectively.
This is because the coordination of the players' strategies
depends on the availability of information 
of the population's responses to $D$ different strings.
When $N\ll D$, the number $sN$ of strategies possessed by the population
is much less than the number $D$ of input states of the strategies.
This makes the coordination of players difficult,
limiting the market efficiency.

On the other hand, the dynamics is highly periodic 
in the low complexity phase \cite{savit}.
This periodicity is caused by the existence of some players 
who, on losing the game, switch their strategies
in an attempt to win at the next occurence of the same game history.
However, their switch turns out to tip the balance
of the strategy distribution,
and the previously winning bit (which used to be the minority bit)
becomes at the next occurence a losing bit
(which newly becomes the majority bit).
This switching can go on back and forth indefinitely,
resulting in the periodic dynamics.
In other words, the game is undesirably influenced by players 
who are maladaptive to the environment.
This causes the very large variances of the decisions,
which becomes larger than that of random decisions
in most of the low complexity phase.

Hence we show, on the same figure,
the variance for different values of diversity $\rho$.
It is observed that the variance decreases significantly
with diversity in the low complexity phase,
although it remains unaffected in the high complexity phase.
Furthermore, for a game efficiency
prescribed by a given value of variance $\sigma^2/N$,
the required complexity of the players is much reduced.

Figure \ref{sd}(a) also illustrates   
the scaling of the variance
with respect to complexity and diversity.
When both $m$ and $N$ vary,
we find that the data points collpase together
for the same values of $\rho$ and $\alpha$.
This means that randomness affects the system behavior
in multiples of $N$.

The scaling of the variance with diversity is further confirmed
in Figure \ref{sd}(b) for given memory sizes $m$.
Furthermore, except for the several points with very small $R$
($R=1, 3, 7$), the variance scales as $\rho^{-1}$
over a wide range of diversity.

To illustrate the physical picture,
we consider the fraction of players switching strategies
at low values of $m$.
In this case, there are relatively few pairs of possible strategies.
Hence when $R$ is not too small, 
the virtual point distribution for a strategy pair can be described
by a Gaussian distribution with standard deviation scaling as $\sqrt R$.
At each time step,
the fraction of players switching strategies
is determined by those whose strategies have equal virtual points.
Hence this fraction scales as $1/\sqrt R$,
leading to a variance of $\sigma^2/N\sim\rho^{-1}$.
(The scaling relation deviates at small $R$
because the multinomial nature of the virtual point distribution
deviates from the Gaussian profile at larger $R$.)
Qualitatively, with random initial conditions,
there is now a diversity of players
having different preferences of strategies.
At each time step,
only those with weak preferences switch strategies.
This greatly reduces the maladaptive behavior and the population variance.

\begin{figure}[ht]
\centerline{\epsfig{figure=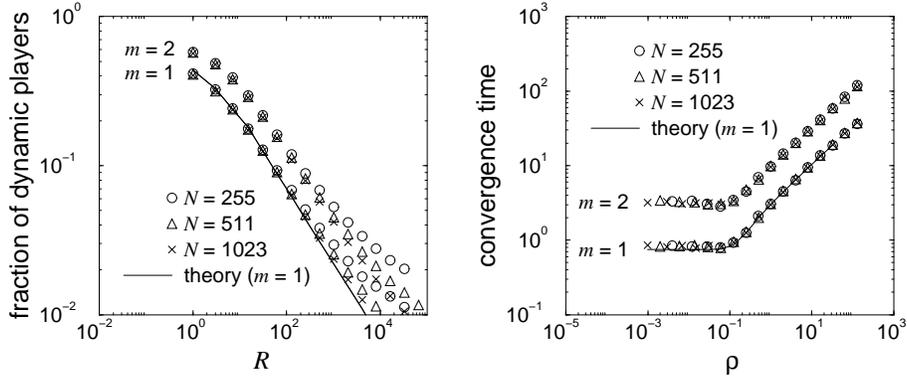,width=0.95\linewidth}}
\caption{\small The dependence of
(a) the fraction of dynamic players on the randomness, and
(b) the convergence time on the diversity, 
at $s=2$ and averaged over 1024 samples for each data point.
}
\label{dynamic}
\end{figure}

This physical picture is further illustrated
by considering the fraction of players
who switches their strategies
even after the game has reached the steady state.
These {\it dynamic players} hold strategy pairs
whose virtual point differences are distributed near zero,
and hence should scale as $1/\sqrt R$.
This is confirmed in Fig. \ref{dynamic}(a).

On the other hand,
since diversity reduces the fraction of players switching strategies
at each time step,
it also slows down the convergence to the steady state.
It has been argued that the dynamics of the game
proceeds in a direction which reduces the variance \cite{challet}. 
Since the step size scales as $1/\sqrt R$,
and the fractional difference of the population with decisions 1's and 0's 
has an initial variance scaling as $1/\sqrt N$,
the convergence time scales as $\rho^{1/2}$.
As shown in Fig. \ref{dynamic}(b),
the dynamics converges almost instantly to the steady state for small $R$.
Beyond that, the predicted scaling relation holds
for various values of $N$ over a wide range of diversity.

\begin{figure}[ht]
\centerline{\epsfig{figure=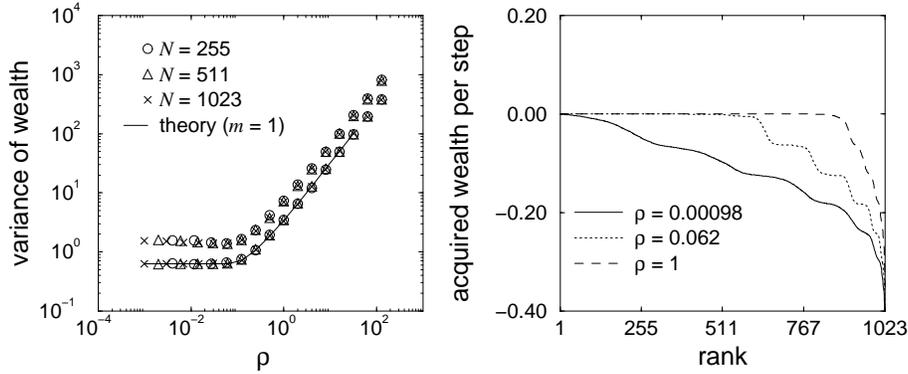,width=0.95\linewidth}}
\caption{\small (a) The dependence of the variance of wealth
among the frozen players;
(b) the individual wealth accquired per step of all players 
(in descending order) for three games with different values of $\rho$
at $m=4$, $s=2$ and $N=1023$, averaged over 1024 samples.
}
\label{wealth}
\end{figure}

It is instructive to consider how the distribution of wealth or resources
acquired by the players changes with diversity,
both at the transient and steady states.
We find that all {\it frozen} players
(that is, players who do not switch their strategies at the steady state)
have constant average wealth at the steady state.
Hence any differences in their wealth
are a consequence of the transient dynamics.
The variance of the wealth distribution should scale
as the square of the convergence time,
that is, as $\rho$.
As shown in Fig. \ref{wealth}(a),
this scaling relation indeed holds
beyond the region of small $R$
where the dynamics converges almost instantly to the steady state.
It is interesting to note the similarities
between Figs. \ref{wealth}(a) and \ref{dynamic}(b),
demonstrating that they have the same origin.

Figure \ref{wealth}(b) shows the individual wealth
acquired per step by the players at the steady state,
ranked from left to right in descending order
for games with various diversity.
For $R=0$, the game is relatively inefficient,
since even the most successful players cannot
break even in their wealth acquisition.
When diversity increases,
there is an increasing fraction of players 
with maximum individual wealth.
It is interesting to note that the individual wealth of the players
has a maximum of 0 per step
in all cases with nonvanishing diversity,
indicating the tendency of the game
to distribute resource evenly.
The collective wealth acquired by the game,
or the overall resource utilization,
is measured by the area under the curve of individual wealth,
which again increases with diversity.

\section{Dynamics of the MG}

We have formulated a theory for the dynamics of the MG
at small memory sizes $m$.
Consider the history of the game denoted by the series $\sigma(t)=0, 1$.
We describe the state of the game at time $t$
by an integer $\mu(t)$ of modulo $D$, where
\begin{equation}
        \mu(t)=\sum_{t'=0}^{m-1}\sigma(t-t')2^{t'}.
\end{equation}
Let $\xi^\mu_\alpha$ be the decision
of strategy $\alpha$ at state $\mu$.
In the following analysis,
we will write $\xi^\mu_\alpha = \pm 1$ for decisions 1 and 0 respectively.
The strategies $\alpha$ are labelled from 1 to $2^D$.
The $D$-dimensional phase space of the game is described
by the collective decision components
\begin{equation}
        A^\mu(t)=\frac{1}{N}\sum_\alpha n_\alpha(t)\xi^\mu_\alpha,
\end{equation}
where $n_\alpha(t)$ denotes the number of players
using strategy $\alpha$ at time $t$.
Note that one of these states corresponds
to the historical state of the game,
and is denoted by $\mu^*(t)$.

Below, we compute $n_\alpha(t)$ for $s=2$;
generalization to other values of $s$ is straightforward.
Let $S_{\alpha\beta}(\omega)$
be the number of players holding strategies $\alpha$ and $\beta$
(with $\alpha<\beta$),
and the virtual point of strategy $\alpha$ is initially displaced
by $\omega$ with respect to $\beta$.
The average of $S_{\alpha\beta}(\omega)$ is given by
\begin{equation}
        \langle S_{\alpha\beta}(\omega)\rangle
        =\frac{N}{2^{2D-1}}\left(
        \begin{matrix}
        R\cr
        (R-\omega)/2
        \end{matrix}\right)
        \frac{1}{2^R}.
\label{sab}
\end{equation}
This equation is valid for general values of $R$.
In particular, when $R$ is not too small,
the binomial distribution in Eq. (\ref{sab})
approaches a Gaussian distribution with variance $R$,
rendering the analysis even more transparent.

The key to analysing the game dynamics with random initial conditions
is the observation that the virtual points of all strategies
displace by exactly the same amount when the game proceeds,
though their initial values may be different.
Hence for a given strategy pair,
the profile of the virtual point distribution remains binomial,
but the peak position shifts with the game dynamics.
If the virtual point displacement of strategy $\alpha$
is $\Omega_\alpha(t)$ at time $t$,
then the players holding strategies $\alpha$ and $\beta$
make decisions according to strategy $\alpha$
if $\omega+\Omega_\alpha(t)-\Omega_\beta(t)>0$,
and strategy $\beta$ otherwise.

Let us consider the change in $A^\mu(t)$
when state $\mu$ is the historical state.
Since the winning state is $-{\rm sgn}A^\mu(t)$, 
the virtual points of strategy $\alpha$ shift
from $\Omega_\alpha(t)$ to $\Omega_\alpha(t)-{\rm sgn}A^\mu(t)\xi^\mu_\alpha$.
Changes in the collective decision are only contributed by players with
virtual points on the verge of switching signs, that is, 
$\omega+\Omega_\alpha(t)-\Omega_\beta(t)=\pm 1$,
and $\xi^\mu_\alpha-\xi^\mu_\beta=\mp 2{\rm sgn}A^\mu(t)$.
Hence we have
\begin{eqnarray}
        &&A^\mu(t+1)-A^\mu(t)
        =-{\rm sgn}A^\mu(t)\frac{2}{N}\sum_{\alpha<\beta}
        \bigl[S_{\alpha\beta}(-1-\Omega_\alpha(t)+\Omega_\beta(t))
        \nonumber\\
        &&\delta(\xi^\mu_\alpha-\xi^\mu_\beta,-2{\rm sgn}A^\mu(t))
        +S_{\alpha\beta}(1-\Omega_\alpha(t)+\Omega_\beta(t))
        \delta(\xi^\mu_\alpha-\xi^\mu_\beta,2{\rm sgn}A^\mu(t))\bigr],
\label{dec}
\end{eqnarray}
where $\delta(m,n)$ denotes the Kronecka delta.
In the region where $D\ll\ln N$, 
we have $S_{\alpha\beta}(\omega)\gg 1$, 
and the decision component in Eq. (\ref{dec}) is self-averaging.
Writing
\begin{equation}
        \delta(\xi^\mu_\alpha-\xi^\mu_\beta,\pm 2)
        =\frac{1}{4}(1-\xi^\mu_\alpha\xi^\mu_\beta
        \pm\xi^\mu_\alpha\mp\xi^\mu_\beta),
\end{equation}
and observing that the magnitude of virtual point displacements
are typically much less than the width $\sqrt R$
of the virtual point distribution,
so that $\langle s_{\alpha\beta}(\omega)\rangle$
can be approximated by its value at $\omega=0$,
we arrive at
\begin{equation}
        A^\mu(t+1)-A^\mu(t)=-{\rm sgn}A^\mu(t)
	\frac{1}{2^{2D}}\sum_{\alpha\beta}
        \sqrt{\frac{2}{\pi R}}(1-\xi^\mu_\alpha\xi^\mu_\beta).
\end{equation}
Since $\sum_\alpha\xi^\mu_\alpha = 0$, the final result is
\begin{equation}
        A^\mu(t+1)-A^\mu(t)=-{\rm sgn}A^\mu(t)\sqrt{\frac{2}{\pi R}},
        \quad
        \mu=\mu^*(t).
\label{hist}
\end{equation}
Similarly, one can verify that
if $\nu$ is not the historical state at time $t$, then
\begin{equation}
        A^\nu(t+1)-A^\nu(t)=0,
        \quad
        \nu\ne\mu^*(t).
\label{nhist}
\end{equation}
Now we can consider the steady state dynamics,
which has a period of $2D$.
This is consistent with the picture that
the dynamics proceeds in the direction
which reduces the variance of the decisions,
as evident in Eq. (\ref{hist}).
Concretely, the state evolution is given by the integer equation
\begin{equation}
        \mu^*(t+1)=\mod(2\mu^*(t)+\sigma(t), D),
\end{equation}
so that every state $\mu$ appears as historical states two times
in a steady-state period,
with $\sigma(t)$ appearing as 0 and 1, each exactly once.
One occurence brings the decision component $A^\mu$
from positive to negative,
and another bringing it back from negative to positive,
thus completing a cycle.
Due to the maladaptive nature of the dynamics,
the component keeps on oscillating,
but never reaches the zero value.
For examples, the steady state for $m=1$ is given by the sequence
$\mu(t) = \sigma(t) = 0, 1, 1, 0$,
where one notes that both states 0 and 1 are followed by 0 and 1 once each.
For $m=2$, there are 2 attractors,
described by  the sequences $\mu(t) = 0, 1, 3, 3, 2, 1, 2, 0$
and $\mu(t) = 0, 1, 2, 1, 3, 3, 2, 0$.
Again, one notes that each of the states 0, 1, 2, 3 are followed
by an even ($\sigma(t)=0$) and an odd state ($\sigma(t)=1$) once each.

As a result, each state is eventually confined in a $D$-dimensional hypercube 
of size $\sqrt{2/\pi R}$,
irrespective of the initial position of the decision components.
This confinement enables us to compute the variance of the decisions.
Without loss of generality,
let us relabel the time steps in the periodic attractor,
with $t=0$ corresponding to the state with $\mu^*(t)=0$
and $\mu^*(t+1)$ being odd,
and let us denote as $t_\mu$ the step at which state $\mu$ first appears
in the relabeled sequence.
(For example, for the first $m=2$ attractor mentioned above,
$t_0=0$, $t_1=1$, $t_2=4$ and $t_3=2$.)
When state $\mu$ first appears in the attractor on or after $t=0$,
its decision component is $A^\mu(t=0)$
by virtue of Eq. (\ref{nhist}),
and the winning state is $2\sigma(t_\mu)-1$.
When state $\mu$ appears in the attractor the second time,
its decision component is $A^\mu(t=0)+[2\sigma(t_\mu)-1]\sqrt{2/\pi R}$
from Eq. (\ref{hist}),
and the winning state is $1-2\sigma(t_\mu)$.
Since the winning state is determined by the minority decision,
these impose the conditions
\begin{equation}
	-\sqrt{\frac{2}{\pi R}}<A^\mu(t=0)[2\sigma(t_\mu)-1]<0.
\end{equation}
Suppose the game starts from the initial state $A^\mu_0$,
which are Gaussian numbers with mean 0 and variance $1/N$.
They change in steps of size $\sqrt{2/\pi R}$
until they reaches the attractor,
whose $2D$ historical states are then given by
\begin{equation}
        \sqrt{\frac{2}{\pi R}}\left\lceil
        \sqrt{\frac{\pi R}{2}}A^\mu_0\right\rceil
        \quad{\rm and}\quad
        \sqrt{\frac{2}{\pi R}}\left\{\left\lceil
        \sqrt{\frac{\pi R}{2}}A^\mu_0\right\rceil-1\right\},
\end{equation}
where $\lceil x\rceil$ represents the decimal part of $x$.
This corresponds to a variance of decisions given by
\begin{equation}
        \frac{\sigma^2}{N}
	\equiv\frac{N}{4}\langle[A^{\mu^*(t)}(t)-
	\langle A^{\mu^*(t)}(t)\rangle]^2\rangle
        =\frac{1}{2\pi\rho}f(\rho),
\label{var}
\end{equation}
where $f(\rho)$ is a smooth function of $\rho$,
which approaches $(1-1/4D)/3$ for $\rho\gg 1$.
Note that Eq. (\ref{var}) holds for general values of $m$,
provided that they are not too large.
This is verified in Fig. \ref{sd}(b)
where the simulation results for $m=1$ and $m=2$ collapse,
and agree with the theory.
Results for higher values of $m$, 
to be presented elsewhere, 
also yield excellent agreement with simulations.
For small values of $R$, 
the $2D$ historical states can be computed similarly, 
taking into account the multinomial nature of the virtual point distribution, 
and yield excellent agreement with simulation results.

Other parameters, such as the convergence time,
the fraction of dynamic players, and the wealth distribution,
can be computed from the same physical picture of the game dynamics.
Detailed derivations will be presented elsewhere.

\section{Conclusion}

We have studied the effects of diversity
in the initial preference of strategies 
on a version of large population games.
We find that it leads to an increase in the system efficiency, 
and a reduction of the required complexity for a given efficiency. 
Both theoretical and simulational studies
show that scaling relations exist for
the dependence of the efficiency on the diversity.
The variance of decisions in the low complexity phase decreases,
showing that the maladaptive behavior is reduced.
Likewise, the resource at the steady state 
is also more evenly distributed.
On the other hand, the convergence time increases with diversity.

Theoretical studies confirm the physical picture 
that the game proceeds in steps 
which tend to reduce the difference 
between the population of the two decisions, 
projected along one state at each step. 
Maladaptation prevents the difference from reducing to zero, 
overshooting at each step 
by a step size scaling as $1/\sqrt R$, 
resulting in the periodic attractor 
at the low complexity phase.
This provides an explanation for the scaling behavior of 
the variance of decisions and other parameters
as a function of the diversity.

The sensitivity of the steady state to the initial conditions
has implications to adaptation and learning in games.
When distributive learning algorithms
of the reinforcement-learning type are devised,
it may be possible that a Nash equilibrium cannot be reached.
In the present example of MG, we find that the dynamic players
lose wealth continuously at the steady state,
because of their untimely switching between two strategies.
In other words, the dynamical rules of virtual point updates
prevent them from adopting the best response in a timely fashion.

Hence care should be taken to avoid maladaptation.
If maladaptation is indeed a problem,
it will be useful to limit its effects
by introducing diversity among the players,
so that the phase space is more efficiently explored.
Like the present experiment,
diversity may help us to attain an increased system efficiency 
with less complex players.

\subsubsection*{Acknowledgments}

We thank Zhuo Gao and Leihan Tang for fruitful discussions.
This work is supported by the research grant HKUST6153/01P
from the Research Grant Council of Hong Kong.


\small{

}

\end{document}